\documentclass[%
twocolumns,
reprint,
 amsmath,amssymb,
 aps,
 amsfonts,
 mathrsfs
pra,
]{revtex4-2}

\usepackage{graphicx}
\usepackage{dcolumn}
\usepackage{bm}

\usepackage{algorithm}
\usepackage{algpseudocode}

\usepackage{xcolor}%
\usepackage{textcomp}%
\usepackage{booktabs}%
\usepackage{braket}

\newtheorem{rem}{Remark}

\begin{document}


\title{Cluster property and Bell's inequalities}


\author{F. Benatti}
\affiliation{Department of Physics, University of Trieste, Trieste, Italy\\
INFN, Trieste, Italy}

\author{R. Floreanini}
\affiliation{INFN, Trieste, Italy}

\author{H. Narnhofer}
\affiliation{Institute of Theoretical Physics, University of Wien, Wien,  Austria}


\begin{abstract}
Among the many loopholes that might be invoked to reconcile local realism with the experimental violations of Bell's inequalities, the space-dependence of 
the correlation functions appears particularly relevant for its connections with the so-called cluster property, one of the basic ingredient of axiomatic quantum field theory.
The property states that the expectation values of products of observables supported within space-like separated space-time regions factorize. Actually, in some massive models the factorization is exponentially fast with respect to the distance between the systems possibly involved in actual experiments.
It is then often argued that considering the space dependence of the quantities involved in the Bell's like inequalities would eventually not violate  them and  thus support the reproducibility of the quantum behaviour by a suitable local hidden variable model.
In this note, we show when this is actually the case and how non-local effects can still be visible.
\end{abstract}


\maketitle

\section{Introduction}

Quantum entanglement plays a peculiar role in local quantum field theory~\cite{Haag,Araki} both in relation to non-local correlations  and to the fate of the Bell's inequalities whose violations should be able to put it into evidence~\cite{SummersWerner1}--\cite{Summers2011}.
A consequence of the Reeh-Schlieder theorem is that the vacuum state is entangled over the operators located within two spatially separated double cones, no matter how far they are apart.  In addition, the expectation value of products of observables localized within the double cones do not in general factorize into the product of the single observable expectation values. This fact actually holds for any state which needs a finite amount of energy to be constructed out of the vacuum~\cite{Halvorson2000,Halvorson2001}.
For this reason, in the following we shall  limit our considerations to the vacuum state.
However, the so-called cluster property, also known as cluster decomposition~\cite{Streater1989,Strocchi2021}, asserts that such a lack of factorization very rapidly fades away with increasing distance between the two of them.

Well known witnesses of quantum entanglement, namely of non-local quantum correlations, are the violations of 
Bell's inequalities~\cite{Bell1964}--\cite{Bertlmann2016}, 
in particular in their (CHSH) spin $1/2$ formulation by Clauser, Horne, Shimony and Holt~\cite{CHSH1969}: these inequalities concern correlations among spin degrees of freedom of two two-level quantum systems independently from how far apart they are.
However, in almost all approaches the spatial degrees of freedom are disregarded, while their impact might be relevant if there is a dynamics that tend to spatially separate the systems involved. Indeed, it is argued~\cite{SummersWerner1,SummersWerner2,Volovich1,Volovich2} that, if the 
space-dependence of quantum states is taken into account, then the cluster property, should dramatically weaken the violations  of the CHSH inequalities, by exponentially damping the correlation functions used to check them.

Yet, plenty of experiments have now confirmed that the CHSH inequalities are indeed violated~\cite{Aspect1982}--\cite{Note}.
In this short note, we purport that these evidences can be reconciled with the cluster property by adapting the spatial part of the measured spin observables  to the dynamics that moves the particles away one from the other, this latter being a procedure actually performed in experiments with massive particles with the detectors being positioned where a sufficiently high statistics can be retrieved.

\section{Cluster property}

In Axiomatic Quantum Field Theory, the standard technical setting is that of a suitable $C^*$ or von Neumann algebra $\mathcal{A}$ of field operators equipped with a positive, normalized linear expectation, namely a state, $\omega:\mathcal{A}\mapsto\mathbb{C}$, that provides all operators in $\mathcal{A}$ with specified mean values. 
As a typical example, the vacuum state vector $\vert\Omega\rangle$, such that $a\vert\Omega\rangle=0$ for any annihilation operator $a$ of the field,  corresponds to an expectation 
$\omega(A)=\langle\Omega\vert A\vert\Omega\rangle$ for all $A\in\mathcal{A}$.
In the algebraic context, the cluster property asserts that, with respect to the vacuum state, the expectation values  
$\omega(A\,B)$ (two-point correlation functions) of two
free-field operators $A$ and $B$ localized in bounded spatial regions factorize asymptotically in time as the support of $A$ is moved sufficiently away from the one of $B$ by the free dynamics 
$\alpha_t:A\mapsto \alpha_t(A)$; namely,
\begin{equation}
\label{Cluster2}
\lim_{t\to\pm\infty}\omega(\alpha_t(A)\,B)=\omega(A)\,\omega(B)\ .
\end{equation}
Perhaps, the simplest concrete example of such a behaviour is when the dynamics $\alpha_t$ is replaced by the space-translations $\beta_{\boldsymbol{y}}$. Namely, consider free fields with  the smeared annihilation operators
$$
a(f)=\int_{\mathbb{R}}{\rm d}\boldsymbol{k}\, \tilde{f}(\boldsymbol{k})\,a_{\boldsymbol{k}}\ ,\quad a_{\boldsymbol{k}}\,a^\dag_{\boldsymbol{q}}\,-
\,a^\dag_{\boldsymbol{q}}\,a_{\boldsymbol{k}}=\delta(\boldsymbol{k}-\boldsymbol{q})\ ,
$$
where $\tilde{f}(\boldsymbol{k})$ is the momentum-space Fourier transform of a smooth, square summable function $f(\boldsymbol{x})\in L^2(\mathbb{R}^3)$ and $a_{\boldsymbol{k}}, a^\dag_{\boldsymbol{k}}$ denote Bosonic annihilation and creation operators of single particle states with momentum $\boldsymbol{k}$. If $V_{\boldsymbol{y}}$, $\boldsymbol{y}\in\mathbb{R}^3$, denotes the unitary space-translation operator on $L^2(\mathbb{R}^3)$ such that
$$
(V_{\boldsymbol{y}}f)(\boldsymbol{x})=f_{\boldsymbol{y}}(\boldsymbol{x}):=f(\boldsymbol{x}+\boldsymbol{y})\ ,\qquad
\tilde{f}_{\boldsymbol{y}}(\boldsymbol{k})={\rm e}^{i\boldsymbol{k}\cdot\boldsymbol{y}}\,\tilde{f}(\boldsymbol{k})\ ,
$$
the space-translations $\beta_{\boldsymbol{y}}$ on $\mathcal{A}$ are implemented by $\beta_{\boldsymbol{y}}(a(f))=a(f_{\boldsymbol{y}})$.
The vacuum state $\vert\Omega\rangle$ is such that $a(f)\vert\Omega\rangle=0$, while $a^\dag(f)$ creates a single-particle field state, $a^\dag(f)\vert\Omega\rangle=\vert f\rangle$.
Thus, one has $\omega(a(f))=0$, for all $f(\boldsymbol{x})\in L^2(\mathbb{R}^3)$, and
$$
\omega(a(f)a^\dag(g))=\langle a^\dag(f)\Omega\vert a^\dag(g)\vert\Omega\rangle=\langle f\vert g\rangle
$$
for all $f(\boldsymbol{x}), g(\boldsymbol{x})\in L^2(\mathbb{R}^3)$.
Then, the Riemann-Lebesgue lemma~\cite{ReedSimon} yields
\begin{eqnarray}
\nonumber
\lim_{\|\boldsymbol{y}\|\to+\infty}\omega(a(f_{\boldsymbol{y}})a^\dag(g))&=&\lim_{\|\boldsymbol{y}\|\to+\infty}\int_{\mathbb{R}^3}{\rm d}\boldsymbol{k}\,{\rm e}^{i\boldsymbol{k}\cdot\boldsymbol{y}}\tilde{f}(\boldsymbol{k})\,\tilde{g}^*(\boldsymbol{k})\\
\label{Cluster3}
&=&0=\omega(a(f))\,\omega(a^\dag(g))\ .
\end{eqnarray}
Such an asymptotic behaviour generalizes the fact that any two initially overlapping compact supports can be made disjoint by a suitable
space-translation, so that 
$$
\omega(a(f_{\boldsymbol{y}})a^\dag(g))=\int_{\mathbb{R}^3}{\rm d}\boldsymbol{x}\
f(\boldsymbol{x}+\boldsymbol{y})\,g^*(\boldsymbol{x})\,=\,0\ ,
$$
as soon as $f_{\boldsymbol{y}}(\boldsymbol{x})g(\boldsymbol{x})=0$,
thus retrieving the cluster property~\eqref{Cluster2} with respect to the space-translation group, assuming $\omega(A)=\omega(B)=0$.

In the relativistic setting, consider a massive scalar, Bosonic free field in $3$ space-dimensions 
\begin{equation}
\label{Cluster4}
\Phi(t,\boldsymbol{x})=\frac{1}{(\sqrt{2\pi})^3}\int_{\mathbb{R}^3}\frac{{\rm d}\boldsymbol{k}}{\sqrt{2k_0}}\,\left({\rm e}^{ikx}\,a^\dag_{\boldsymbol{k}}+\,{\rm e}^{-ikx}\,a_{\boldsymbol{k}}\right)\ ,
\end{equation}
where $kx=k_0\,t-\boldsymbol{k}\cdot\boldsymbol{x}$ and $k_0=\sqrt{\boldsymbol{k}^2+m^2}$ 
and $a^\dag_{\boldsymbol{k}}$ acting on the vacuum state $\vert\Omega\rangle$ creates a particle of mass $m$ and momentum $\boldsymbol{k}$.
Notice that $\Phi(t,\boldsymbol{x})=\alpha_t\beta_{\boldsymbol{x}}(\Phi(0,\boldsymbol{0}))$, where
$\alpha_t(a(f))=a(f_t)$ and $f_t$ solves the Klein-Gordon equation $(\partial^2_t -\nabla^2_{\boldsymbol{x}}+m^2)f_t(x)=0$.

For space-like separations and 
with respect to the space-translation group, the equal time, two-point correlation functions $\omega(\Phi(t,\boldsymbol{x})\Phi(t,\boldsymbol{y}))$ explicitly
read~\cite{Casalbuoni}
\begin{equation}
\label{Cluster5}
\omega(\Phi(t,x)\Phi(t,y))=-\frac{m}{8\pi}\frac{H^{(1)}_1(im\|\boldsymbol{x}-\boldsymbol{y}\|)}{\|\boldsymbol{x}-\boldsymbol{y}\|}\ ,
\end{equation}
where $H^{(1)}_1$ is a Hankel function of order $1$. Its asymptotic behaviour is, for large $\|\boldsymbol{x}-\boldsymbol{y}\|$, given by
\begin{equation}
\label{Cluster6}
H^{(1)}_1(im\|\boldsymbol{x}-\boldsymbol{y}\|)\simeq -\sqrt{\frac{2}{m\pi\|\boldsymbol{x}-\boldsymbol{y}\|}}\, {\rm e}^{-m\|\boldsymbol{x}-\boldsymbol{y}\|}\ .
\end{equation}
Therefore, an exponential decay of the spatial correlation functions is retrieved  for space-distances larger than the Compton wavelength $\lambda=1/m$; notice that for an electron $\lambda$ is of the order of $10^{-12}$ $m$.

Notice that expectation values of generic products of fields reduce to sums of products of the two-point function as in~\eqref{Cluster5}; they therefore
inherit the cluster property~\eqref{Cluster2}.

\begin{rem}
\label{rem1}
If self-adjoint, the field operators $A$ and $B$ represent physical observables: these can be simultaneously measured when they commute, as is the case when their supports are space-like separated.
Therefore, the cluster property as expressed by~\eqref{Cluster2}, reflects  the statistical independence, respectively asymptotic statistical independence, of commuting $A$ and $B$ with respect to the reference state, the vacuum in the specific case.
Particularly when dealing with Bell's inequalities,
one is interested in time-correlation functions of the form
$\langle\psi\vert\alpha_t(A)\,B\,\vert\psi\rangle$. Namely, not  with respect to the vacuum state $\vert\Omega\rangle$ as in~\eqref{Cluster2}, rather with respect to generic vector states $\vert\psi\rangle$ of the field particles. 
The factorization of the vacuum correlation functions extends in the following way.  
The state $\vert\psi\rangle$ can be reached within any fixed accuracy by acting on the vacuum, $\vert\psi\rangle=P\vert\Omega\rangle$, with a polynomial 
$P$ in creation operators localized in a spatial region $V_P$.
Then, 
$\langle\psi\vert\, \alpha_t(A)\,B\vert\psi\rangle=\langle\Omega\vert P^\dag \alpha_t(A)\,B\,P\vert\Omega\rangle$.
Physical observables $A$ and $B$ can always be thought localized within a bounded spatial region $V_{AB}$; if, for sufficiently large times $t$, the dynamics localizes $\alpha_t(A)$ in a region causally disconnected from the bounded region $V_P\bigcup V_{AB}$, then $[\alpha_t(A)\,,\,B\,P]=0$ so that:
\begin{eqnarray}
\nonumber
&&\hskip-.5cm
\langle\psi\vert \alpha_t(A)\,B\vert\psi\rangle=\langle\Omega\vert P^\dag\,\alpha_t(A)\,B\,P\vert\Omega\rangle\\
\nonumber
&&\hskip .5cm
=\langle\Omega\vert P^\dag\,B\,P\alpha_t(A)\vert\Omega\rangle\mapsto\langle\Omega\vert P^\dag\,B\,P\vert\Omega\rangle\,\omega(A)\\
&&
\hskip 1cm
=\langle\psi\vert B\vert\psi\rangle\,\omega(A)\ ,
\label{Cluster8}
\end{eqnarray}
where the clustering property~\eqref{Cluster2} and the time-invariance of the state, $\omega(\alpha_t(A))=\omega(A)$, have been used in the second line; indeed, as remarked before, the cluster property~\eqref{Cluster2} holds for any finite-energy state created from the vacuum.
Such an argument can be further extended to time-invariant reference states other than the vacuum, for instance to thermal states, by means of the so-called modular theory~\cite{Summers1990}--\cite{Halvorson2001}.
\end{rem}

\section{Bell's inequalities: CHSH formulation}

Consider two spin $1/2$ particles, the single system Pauli matrices $\sigma_{x,y,z}\otimes \mathbb{I}$, $\mathbb{I}\otimes\sigma_{x,y,z}$ and local observables of the form
$(\boldsymbol{\sigma}\cdot\mathbf{a})\otimes(\boldsymbol{\sigma}\cdot\mathbf{b})$,where 
$\boldsymbol{\sigma}=(\sigma_x,\sigma_y,\sigma_z)$,
with $\boldsymbol{a}$ and $\boldsymbol{b}$ unit vectors in $\mathbb{R}^3$.

Denoting by $\vert 0\rangle$ and $\vert 1\rangle$ the eigenstates of $\sigma_z$ and by $\vert ij\rangle$ the tensor products $\vert i\rangle\otimes \vert j\rangle$,
the projector onto the singlet state $\vert\Sigma\rangle=(\vert01\rangle-\vert10\rangle)/\sqrt{2}$  can be recast as
\begin{equation}
\label{singlet}
\vert\Sigma\rangle\langle\Sigma\vert=\frac{1}{4}\Big(\mathbb{I}\otimes\mathbb{I}-
\sigma_x\otimes\sigma_x-\sigma_y\otimes\sigma_y-\sigma_z\otimes\sigma_z\Big)
\ .
\end{equation}
Since the Pauli matrices are traceless and
$\displaystyle(\boldsymbol{\sigma}\cdot\boldsymbol{a})(\boldsymbol{\sigma}\cdot\boldsymbol{b})=\boldsymbol{a}\cdot\boldsymbol{b}+i(\boldsymbol{a}\times\boldsymbol{b})\cdot\boldsymbol{\sigma}$,
one computes
\begin{equation}
\label{mvsinglet}
\langle(\boldsymbol{\sigma}\cdot\boldsymbol{a})\otimes(\boldsymbol{\sigma}\cdot\boldsymbol{b})\rangle_\Sigma=
{\rm Tr}\Big(\vert\Sigma\rangle\langle\Sigma\vert\,(\boldsymbol{\sigma}\cdot\boldsymbol{a})\otimes(\boldsymbol{\sigma}\cdot\boldsymbol{b})\Big)=-\boldsymbol{a}\cdot\boldsymbol{b}\ .
\end{equation}
In a local hidden variable  ($\lambda$) theory, one assumes that the above expectations values can be calculated as mean values  of products of classical stochastic variables $a(\lambda)$ and $b(\lambda)$ with respect to a classical probability distribution $p(\lambda)$, the classical observables assuming only the values $\pm1$ as the quantum ones, $(\boldsymbol{\sigma}\cdot\boldsymbol{a})$ and $(\boldsymbol{\sigma}\cdot\boldsymbol{b})$. Namely,
\begin{equation}
\label{HV1}
\langle(\boldsymbol{\sigma}\cdot\boldsymbol{a})\otimes(\boldsymbol{\sigma}\cdot\boldsymbol{b})\rangle_\Sigma=\int{\rm d}\lambda\, p(\lambda)\,a(\lambda)\,b(\lambda)\ .
\end{equation}
Given spatial directions $\boldsymbol{a}_{1,2}$ and $\boldsymbol{b}_{1,2}$, since $a_{1,2}(\lambda)=\pm1$ and $b_{1,2}(\lambda)=\pm1$, the following upper bound holds, known as the CHSH- form of the Bell's inequalities~\cite{CHSH1969}:
\begin{eqnarray}
\nonumber
&&
\left\vert\int{\rm d}\lambda\, p(\lambda)\,\Big(a_1(\lambda)\,(b_1(\lambda)+b_2(\lambda))\right.\\
&&\hskip 1cm 
\left.+a_2(\lambda)(b_1(\lambda)-b_2(\lambda)\Big)\right\vert\,\leq 2\ ,
\label{HV2}
\end{eqnarray}
while $\boldsymbol{a}_1=(0,0,1)$, $\boldsymbol{a}_2=(1,0,0)$, $\boldsymbol{b}_1=-\frac{1}{\sqrt{2}}(1,0,1)$ and $\boldsymbol{b}_2=\frac{1}{\sqrt{2}}(1,0,-1)$ 
yield
\begin{eqnarray}
\nonumber
&&
\left\langle(\boldsymbol{\sigma}\cdot\boldsymbol{a}_1)\otimes(\boldsymbol{\sigma}\cdot(\boldsymbol{b}_1+\boldsymbol{b}_2)+
(\boldsymbol{\sigma}\cdot\boldsymbol{a}_2)\otimes(\boldsymbol{\sigma}\cdot(\boldsymbol{b}_1-\boldsymbol{b}_2))\right\rangle_\Sigma\\
\label{HV3}
&&\hskip 1cm=2\sqrt{2}\ .
\end{eqnarray}
Violations of the classical upper bound $\leq 2$ in the CHSH inequalities have been reported in quite a number of different experimental 
contexts~\cite{Aspect1982}--\cite{Genovese2005} sophisticated enough to close all possible 
loopholes~\cite{Hensen2015}--\cite{Rosenfeld2017}, the cluster property briefly sketched above apparently offering a challenging one.

\subsection{Cluster property vs CHSH inequalities}
\label{app:simple_case}

Since violating the CHSH inequalities can only be observed from entanglement persisting over long distances, whereas the cluster property states that correlations tend to factorize, a natural suspicion arises: can the latter phenomenon blur the observability of such violations?
As rightly observed in~\cite{Volovich1}, possible problems appear when one introduces the spatial degrees that any quantum system possesses together with its spin degrees of freedom.
Indeed, one needs then consider state vectors $\vert\Psi\rangle$ that are sums of tensor products of a spatial component $\vert\psi\rangle\in L^2(\mathbb{R}^3)$ and a spin
component $\vert\Sigma\rangle$: $\vert\Psi\rangle=\vert\psi\rangle\otimes\vert\Sigma\rangle$. At the same time the observables intervening in the Bell's inequalities~\eqref{HV2} become of the form 
$\displaystyle A_1A_2\otimes((\boldsymbol{\sigma}\cdot\boldsymbol{a})\otimes(\boldsymbol{\sigma}\cdot\boldsymbol{b}))$,
where $A_{1,2}$ are observables physically associated with the spatial locations of the apparatuses acting on the spin degrees of freedom.
Even if such degrees of freedom are dynamically inert, one must nevertheless take into account the dynamics of the spatial degrees of freedom; then, the relevant mean values to be inserted in the CHSH inequalities read
\begin{eqnarray}
\nonumber
&&
\langle\psi_t\otimes\Sigma\vert A_1A_2\otimes((\boldsymbol{\sigma}\cdot\boldsymbol{a})\otimes(\boldsymbol{\sigma}\cdot\boldsymbol{b}))\vert\psi_t\otimes \Sigma\rangle\\
\label{clvsB2}
&&\hskip.3cm
=\omega(C^\dag\alpha_t(A_1A_2)C)\,\langle\Sigma\vert((\boldsymbol{\sigma}\cdot\boldsymbol{a})\otimes(\boldsymbol{\sigma}\cdot\boldsymbol{b}))\vert\Sigma\rangle\ .
\end{eqnarray}
In the above expression we have used the setting of Remark~\ref{rem1}
where the time-evolution of the states $\vert\psi\rangle\mapsto \vert\psi_t\rangle$  has been converted into the Heisenberg dynamics of operators $A_1A_2\mapsto\alpha_t(A_1A_2)$.

Suppose $\omega(A_1A_2)=0$; then, a behaviour of the spatial correlation functions as in~\eqref{Cluster8} would hamper the visibility of possibile violations of the CHSH inequality, especially when the damping is as strong as the exponential one in~\eqref{Cluster6}, no matter the amount of entanglement in the spin degrees of freedom.

For instance, as observed before Remark~\ref{rem1}, in the case of a free scalar particle of the mass of an electron already over distances of few
picometers, violations of the CHSH inequalities would become invisible.

\section{Cluster property and violations of the Bell's inequalities reconciled}

The issue brought to the fore by the cluster property is different form the Bell-locality loophole~\cite{Brunner2014,Scarani2019}. It can be summarized by the following question: \textit{Why violations of the Bell's inequalities and entanglement over large distances~\cite{Ma2012}--\cite{Krutyanskiy2023} can then be observed despite the decay of spatial-correlation functions predicted by the cluster property?}

When the spatial component $\vert\psi\rangle$ in $\vert\Psi\rangle=\vert\psi\rangle\otimes\vert\Sigma\rangle$ changes in time ,
$\vert\psi\rangle\mapsto\vert\psi_t\rangle$, the observables $A_{1,2}$ obviously need be adjusted to the locations of the two quantum systems described by $\vert\psi_t\rangle$; indeed, this is what is practically done in an experimental setting. For instance, if $\vert\psi\rangle=\vert\psi_1\rangle\otimes\vert\psi_2\rangle$, where $\vert\psi_{1,2}\rangle$ are the spatial states of particles $1$ and $2$ moving away from each other, then, at time $t$, the observables $A_1$ and $A_2$ should be localized where the probability of finding particle $1$, respectively particle $2$ is highest.
Of course, this adaptation makes the observables $A_{1,2}$ time- (or space-) dependent, thus possibly counteracting the damping effects of the cluster property as no expectation $\omega(A_1A_2)=0$ can then be extracted, by means of the cluster property,  from $\omega(C^\dag\alpha_t(A_1A_2)C)$.

The case of the space-translations serves as a simple illustration of the mechanism: suppose particle $1$ and particle $2$ are initially supported around the origin
$\boldsymbol{0}$ in space and that particle $1$ is translated to the space location $\boldsymbol{y}$ by the unitary $V_{\boldsymbol{y}}$, while  particle $2$ moves in the opposite direction to the space location $-\boldsymbol{y}$ due to the unitary $V_{-\boldsymbol{y}}$. If the observables $A_{1,2}$ are also located around $\boldsymbol{0}$ and remain such, the mean-value $\omega(C^\dag\alpha_t(A_1A_2)C)$ in~\eqref{clvsB2} reduces to
\begin{eqnarray}
\nonumber
&&
\omega(C^\dag(\beta_{\boldsymbol{y}}(A_1)\beta_{-\boldsymbol{y}}(A_2))C)\\
\label{ClBellreconc1}
&&\hskip.5cm
=\langle\psi_1\vert V^\dag_{\boldsymbol{y}}\,A_1\,V_{\boldsymbol{y}}\vert\psi_1\rangle\,
\langle\psi_2\vert V^\dag_{-\boldsymbol{y}}\,A_2\,V_{-\boldsymbol{y}}\vert\psi_2\rangle\ ,
\end{eqnarray}
so that the vanishing of such a correlation function ensues from the Riemann-Lebesgue lemma.
However, the  decay trivially disappears if one adjusts the locations of  $A_{1,2}$ by shifting them in space, namely by sending $A_1$ into
$V_{\boldsymbol{y}}\,A_1\,V^\dag_{\boldsymbol{y}}$ and $A_2$ into $V_{-\boldsymbol{y}}\,A_2\,V^\dag_{-\boldsymbol{y}}$.
Indeed, in experiments  the location of the measurement apparatuses is adjusted to the space regions where the two particles are driven by the dynamics with highest probability.

\subsection{A non relativistic example}
\label{nrelex}

We now analytically illustrate how the exponential decay of space-time correlation functions can be countermanded by spatially shifting
the spatially located observables according to the space-translations of the states of a bipartite quantum system. We do it by considering two
non relativistic free particles in one-dimension and observe the limitations in the dependence of the initial state and in the dependence of the mass of the particles and its decreasing in time.
As initial bipartite state we take the tensor product of two spatial Gaussians centered at $x=0$ of same width $\sigma$ and mean momentum $\pm p_0$:
\begin{equation}
\label{ex1}
G_{\pm p_0}(x)=\frac{1}{\sqrt[4]{\pi\sigma^2}}\,\exp\left(-\frac{x^2}{2\sigma^2}\,\pm\frac{i}{\hbar} p_0 x\right)\ .
\end{equation}
with Fourier transforms in momentum space that are Gaussian functions centered at $\pm p_0$,
\begin{equation}
\label{ex2}
\widetilde{G}_{\pm p_0}(p)=\sqrt[4]{\frac{\sigma^2}{\pi\hbar^2}}\,\exp\left(-\frac{\sigma^2}{2\hbar^2}\Big(p\mp p_0\Big)^2\right)\ .
\end{equation}
As spatial observables $A_{1,2}$, we take the projectors $\vert\phi_{\pm\eta}\rangle\langle\phi_{\pm\eta}\vert$ onto two Gaussian wave-packets of width $ \Delta$ and zero mean momentum centered at positions $\pm\eta$:
\begin{equation}
\label{ex3}
\phi_{\pm\eta}(x)=\frac{1}{\sqrt[4]{\pi\Delta^2}}\,\exp\left(-\frac{(x\pm\eta)^2}{2\sigma^2}\right)\ .
\end{equation}
Then, $A_1A_2=\vert\phi_{\eta}\rangle\langle\phi_\eta\vert\otimes\vert\phi_{-\eta}\rangle\langle\phi_{-\eta}\vert$ and the mean-value $\omega(C^\dag\alpha_t(A_1A_2)C)$ in~\eqref{clvsB2} concretely reads
\begin{equation}
\label{ex4}
\omega(C^\dag\alpha_t(A_1A_2)C)=\left|\langle U_tG_{p_0}\vert\phi_\eta\rangle\right|^2\,\left|\langle U_tG_{-p_0}\vert\phi_{-\eta}\rangle\right|^2\
\end{equation}
where $U_t$ is the unitary dynamics generated by the free Hamiltonian $\displaystyle H=\frac{p^2}{2m}$.

The states evolve into Gaussians $G_{\pm p_0}(x,t)$ that, at time $t$, are located around the positions
$\displaystyle\pm \frac{p_0}{m}\,t$; in computing the moduli of the scalar products in~\eqref{ex4}, all phases that do not depend on $x$ can be neglected, so that
\begin{eqnarray*}
&&
G_{\pm p_0}(x,t)=(U_tG_{\pm p_0})(x)\\
&&=\frac{1}{\sqrt[4]{\pi\sigma_t^2}}\,\exp\left(-\frac{\Big(x\mp\frac{p_0}{m}t\Big)^2}{2\sigma_t^2}\,
\left(1-i\frac{\hbar t}{m\sigma^2}\right)\,\pm \frac{i}{\hbar}\,p_0\,x\right)\ ,
\end{eqnarray*}
where $\displaystyle\sigma^2_t=\sigma^2+\frac{\hbar^2\,t^2}{m^2\,\sigma^2}$.

Straightforward Gaussian integration then gives
\begin{eqnarray}
\nonumber
&&
\left|\langle U_tG_{\pm p_0}\vert\phi_{\pm\eta}\rangle\right|^2=\frac{2\Delta\sigma_t}{L^2_t}\,
\exp\left(-p^2_0\frac{\Delta^4\sigma_t^2}{\hbar^2L^4_t}\right)\times\\
\nonumber
&&\hskip .5cm
\times
\exp\left(-\Big(\eta-\frac{p_0}{m}t\Big)^2\frac{\Delta^2+\sigma_t^2}{L^4_t}\right)\ \times\\
\label{ex7}
&&\hskip.5cm
\times
\exp\left(-\frac{\Delta^2}{L^4_t}\left(\frac{p_0}{\hbar}\sigma^2+\frac{\hbar t}{m\sigma^2}\eta\right)^2\right)\ .
\end{eqnarray}
Let us choose $\Delta=\sigma$ and $t$ such that $(\hbar\,t)/m\ll\sigma^2$ so that $\sigma_t\simeq\sigma$,
\begin{equation}
\label{ex8}
L_t=\sqrt[4]{(\sigma_t^2+\Delta^2)^2+\frac{\hbar^2t^2}{m^2}\left(\frac{\Delta}{\sigma}\right)^4}\simeq\sqrt{2}\,\sigma\ .
\end{equation}
and
\begin{eqnarray}
\nonumber
\left|\langle U_tG_{\pm p_0}\vert\phi_{\pm\eta}\rangle\right|^2&\simeq&\exp\left(-\frac{\sigma^2}{2\hbar^2}p_0^2\right)\\
\label{ex9}
&\times&
\exp\left(
-\frac{1}{2\sigma^2}\left(\eta-\frac{p_0}{m}t\right)^2\right)\ .
\end{eqnarray}
If $p_0\simeq\hbar/\sigma$, then $(p_0\, t)/(m\sigma)\simeq(\hbar\,t)/(m\,\sigma^2)\ll1$ implies the following exponential damping for $\eta\geq \sigma$:
\begin{equation}
\label{ex10}
\left|\langle U_tG_{\pm p_0}\vert\phi_{\pm\eta}\rangle\right|^2\simeq\frac{1}{\sqrt{e}}\,\exp\left(-\frac{1}{2}\frac{\eta^2}{\sigma^2}\right)\ .
\end{equation}
The damping instead disappears if $\eta$ is chosen to be equal to the center of the particle Gaussian state at time $t$: $\eta=(p_0\,t)/m$.

For large times
$L_t\simeq\sigma_t\simeq(\hbar t)/(m\sigma)$;
therefore, with $\eta$ fixed and increasing time,
\begin{equation}
\label{ex11}
\left|\langle U_tG_{\pm p_0}\vert\phi_{\pm\eta}\rangle\right|^2\simeq \frac{2\Delta m\sigma}{\hbar t}\,\exp\left(-p_0^2\frac{\sigma^2}{\hbar^2}\right)\ .
\end{equation}
Instead, by setting $\eta=(p_0 t)/m$, one retrieves
\begin{eqnarray}
\nonumber
\left|\langle U_tG_{\pm p_0}\vert\phi_{\pm\eta}\rangle\right|^2&=&\frac{2\Delta\sigma_t}{L^2_t}\,
\exp\left(-p^2_0\frac{\Delta^2\sigma_t^2(\Delta^2+\sigma_t^2)}{\hbar^2L^4_t}\right)\\
\label{ex12}
&\simeq&
\, \frac{2\Delta m\sigma}{\hbar t}\,\exp\left(-p_0^2\frac{\Delta^2}{\hbar^2}\right)\ .
\end{eqnarray}
Then, by sharply locating the observables with respect to the initial spatial Gaussian width, $\Delta\ll\sigma$, and adjusting their positions at time $t$, one mitigates the effects of the exponential damping factor in~\eqref{ex7}, leaving only the $1/t$ behaviour associated with the spatial spreading of the wave-packet (in higher dimensions this is replaced by $1/t^d$).
Notice that the overall decrease of the correlation functions is then of the order $1/t^2$ ($1/t^{2d}$ in higher dimensions)). This can be compared with  the one observed in optical fiber technologies~\cite{Zhang2023,Volovich2}, where relativity helps one to connect time and space translation for photons.
Experimentally, the $1/t^2$ decrease is indeed cured by increasing the data statistics.

\section{Conclusions}

We have taken up the issue of  possible contradictions between the violations of the CHSH form of the Bell's inequalities and the cluster property of axiomatic quantum field theory predicting asymptotic factorization of correlation functions of causally disconnected field observables.
The apparent conflict with experimental observations follows from considering not only the spin degrees of freedom of the particles involved in the experiments, but also their almost always neglected spatial behaviour.  However, the damping effects related to the cluster property can be counteracted 
by adapting the spatial location of the measured observables to the spatial displacement  of the particles caused by their dynamics. Nevertheless, the larger the spatial separation, the greater the amount of needed experimental data  might become in order to make a violation of the Bell's inequality visible.

\section*{Acknowledgments}
FB acknowledges financial support from the PNRR MUR project PE0000023-NQSTI.

\end{document}